\documentstyle[11pt,newpasp,twoside,epsf]{article}
\def\emphasize#1{{\sl#1\/}}

\def\edcomment#1{\iffalse\marginpar{\raggedright\sl#1\/}\else\relax\fi}
\reversemarginpar

\begin{document}
\title{Elliptical Galaxies: Darkly Cloaked or Scantily Clad?}
\author{A. J. Romanowsky$^1$, 
N. G. Douglas$^2$, 
K. Kuijken$^{2,3}$,
M. R. Merrifield$^1$,\\
M. Arnaboldi$^4$,
N. R. Napolitano$^2$,
H. Merrett$^1$,
M. Capaccioli$^5$,\\
K. C. Freeman$^6$,
O. Gerhard$^7$}
\affil{
$^1$School of Physics and Astronomy, University of Nottingham;
$^2$Kapteyn Institute, Groningen;
$^3$Leiden Observatory;
$^4$INAF--Osservatorio Astronomico di Pino Torinese, Turin; 
$^5$INAF--Osservatorio Astronomico di Capodimonte, Naples;
$^6$RSAA, Mt. Stromlo and Siding Spring Observatories;
$^7$Astronomisches Institut, Universit\"{a}t Basel}

\begin{abstract}
Planetary nebulae (PNe) may be the most promising tracers in the halos of early-type galaxies.
We have used multi-object spectrographs on the WHT and the VLT,
and the new Planetary Nebula Spectrograph on the WHT,
to obtain hundreds of PN velocities in a small sample of nearby galaxies.
These ellipticals show weak halo rotation, which may be consistent with
ab initio models of galaxy formation, 
but not with
more detailed major merger simulations.
The galaxies near $L^*$ show evidence of a universal \emphasize{declining}
velocity dispersion profile,
and
dynamical models indicate the presence of little dark matter within 5~$R_{\rm eff}$---implying
halos either not as massive or not as centrally concentrated as CDM predicts.
\end{abstract}

\section{Introduction}

Dark matter has long been inferred around spiral galaxies
from their flat HI rotation curves;
however, in early-type galaxies (ellipticals and S0s) we do not have this luxury,
and progress in finding their total masses has been slow.
So not only has a fundamental component of the CDM
paradigm remained largely unverified---that
there should be similarly extended, massive dark halos around ellipticals---but
predictions about the detailed halo properties have not been testable
(cf.\ the halo core issues in late-type galaxies discussed
at length in this volume).
There is actually an advantage to studying elliptical halos:
if one can observe at comparable physical radii, the compact nature of ellipticals
implies that this is in a more dark-matter dominated regime than in spirals.
Thus messy baryonic physics should have been less important; and it may be easier
to disentangle the
luminous and dark components'
relative mass contributions.

The nominal tracer in elliptical halos is their \emphasize{integrated stellar light}.
To adequately constrain their dynamics,
kinematical measurements must be of sufficient quality to obtain such higher-order moments
as the Gauss-Hermite $h_4$. But the drop-off in surface brightness
makes this approach nonviable outside an elliptical's central parts.
The best survey so far is of 21 bright, round ellipticals
(see O. Gerhard, this volume),
which found the circular velocity profiles
$v_c(r)$ to be roughly constant out to 1--2~$R_{\rm eff}$ (5--10 kpc),
and ruled out a constant mass-to-light ratio ($M/L$) for
3 of the galaxies.

Alternative probes of elliptical halos include
\emphasize{globular clusters} (GCs), which are handy as bright objects 
spread out to larger radii than the galaxy light (C\^{o}t\'{e} et al. 2001, 2003)---although
they are a disjoint population with different properties to the galaxy.
\emphasize{X-ray emission} from thermalized hot gas filling the halo potential
is also useful (D. Buote, this volume); 
but because the total emission correlates strongly with optical luminosity
($L_{\rm X} \propto L_B^{2-3}$; O'Sullivan et al. 2003), and only the brightest sources
are within easy reach of X-ray telescopes,
the findings are biased toward the more massive systems.
Similarly, \emphasize{strong gravitational lenses} can also be used to probe
into galaxy halos (P. Schneider, this volume), 
but any galaxies with massive, centrally-concentrated halos 
are systematically most likely to be detected as lenses.
\emphasize{Rings and disks} of HI and H$\alpha$-emitting gas are also sometimes found
at large radii (M. Arnaboldi et al., this volume); but these are rare, and
they may not trace a typical population of ellipticals.
\emphasize{Satellite kinematics} (e.g., Prada et al. 2003)
and \emphasize{weak gravitational lensing} (e.g., H. Hoekstra, this volume)
can probe the outermost parts of galaxies,
but their constraints are statistical: they provide
limited information about mass variations with radius and with other galaxy properties.

\emphasize{Planetary nebulae} (PNe)
are arguably the ideal probes because of their simple connection to the
main stellar population of the galaxy (e.g., Peng et al. 2002).
Also, they are not affected by any mass-dust degeneracy (Baes \& Dejonghe 2001),
and their 5007~\AA{} emission lines readily provide precise velocities.

Various studies with the above methods have
found dark matter around individual elliptical galaxies,
but as discussed, the selection effects may be severe.
Weak lensing and satellite studies also imply massive halos 
around typical $L^*$ ellipticals, but further cross-checks are needed.
Indeed, dynamical studies have suggested that
some galaxies are much less dark matter-dominated than others
(Bertin et al. 1994; Gerhard et al. 2001; N. Napolitano et al., this volume).

Besides searching for dark matter,
by probing into elliptical halos we can also test other key properties
against predictions of galaxy formation models.
These include angular momentum and the distribution of stellar and GC orbits.
In the rest of this paper, we present PN kinematical studies of five elliptical
galaxies, and some implications for galaxy structure and formation.

\section{PN Kinematics of M87 and M49: Virgo siblings}

The two brightest ellipticals in the Virgo Cluster,
M87 (=NGC 4486) and M49 (=NGC 4472), 
had their extensive GC systems studied kinematically
(e.g., C\^{o}t\'{e} et al. 2001, 2003).
Both these and X-ray studies
(Schindler et al. 1999; Matsushita et al. 2002)
found massive dark halos.
Yet no stellar nor PN kinematical studies have been done 
in these galaxies' halos for comparison.

We have pursued kinematic follow-up studies of
the large PN samples discovered around these two galaxies with
narrow-band imaging surveys
(Ciardullo et al. 1998; X. Hui et al., unpublished).
PN kinematic studies have till now been sporadically fruitful
because of the low efficiency of spectroscopic follow-up,
which seems to be largely
due to astrometric difficulties.
We used the WYFFOS/AF2 multi-fiber spectrograph on the 4.2-m William Herschel
Telescope (WHT) on La Palma in May 2000 to observe PNe in M49,
giving special attention to
the astrometric solution.
Even with the 2\farcs7 diameter fibers, it appeared that our recovery rate
was not optimal, and we found velocities for $\sim$25 objects.

As a solution to these difficulties,
we devised a new technique, \emphasize{masked counter-dispersed imaging},
which is a combination of traditional multi-slit spectroscopy 
and slitless counter-dispersed imaging (see next section).
Using the FORS2+MXU spectrograph at the ESO 8-m UT2 telescope in April 2001,
we cut standard masks but with the slit sizes increased to 4\arcsec$\times$4\arcsec,
ensuring object recovery.
We took two complementary images of each field with the dispersion direction
reversed, allowing us to cancel out the astrometric errors and derive the
velocities to within $\sim$~15~km~s$^{-1}$.
Completing this program for M87, we found $\sim$~200 PN velocities;
for M49, clouds limited our return to $\sim$~100 PNe.

\section{PN Spectrograph survey: first results}

To overcome the obstacles to observing extragalactic PN kinematics,
our team commissioned a special-purpose instrument,
the \emphasize{Planetary Nebula Spectrograph} (PN.S)
at the WHT (Douglas et al. 2002 and this volume).
Using a slitless spectroscopy technique termed \emphasize{counter-dispersed imaging},
the efficiency of the PN.S has resulted in a breakthrough in this field:
hundreds of PN velocities can be obtained in a single night's observing of a galaxy at a distance of 15~Mpc.

The primary program of the PN.S is to survey a dozen bright ($m_B\leq$~12), round (E0--E2), nearby ($D\la20$~Mpc) ellipticals,
obtaining 100--400 PN velocities in each.
Based on the statistical distribution of galaxy shapes
(Lambas et al. 1992),
we estimate that by selecting for projected axis ratios $q\ga0.8$, 75\% of our sample 
will have intrinsic gravitational potential ellipticities $\epsilon_\Phi \la 0.1$
and thus can be well characterized by spherical dynamical models.
Other than the constraints above, our sample is designed to encompass a broad spectrum
of elliptical galaxies, 
with a large range of stellar light parameters (luminosity, concentration, shape),
rotational importance, and environment.

Our program has so far been beset by bad weather,
but we have obtained extensive data on the galaxies
NGC~821, NGC~3379, and NGC~4494.
For each we have
an initial data set of $\sim$~100 PN velocities out to $\sim$~5~$R_{\rm eff}$,
but with additional data reduction we expect these to increase to $\sim$~200 each.

\section{Angular Momentum: weak rotation}

With substantial PN data sets for five galaxies, we can
first look at their rotational characteristics.
It has long been apparent that the central parts of ellipticals are
much less rotationally-dominated than are spirals (e.g., Fall 1983),
but there may be large amounts of angular momentum
stored in ellipticals' unobserved outer parts---perhaps
even more than in 
spirals because of more dominant major merger histories (J. Primack, this volume).
Such high outer rotation is seen in the GCs around M87,
but it's not clear that these 
trace the properties of the dark matter
and the main stellar population.

None of the galaxies shows a strong increase in rotation with radius
(note though that our sample roundness criterion may introduce a bias toward low spins).
Within $\sim$~3--5~$R_{\rm eff}$,
they have spin parameters $\lambda\arcmin \sim$~0.03--0.07,
and at these radii rotation appears dynamically unimportant:
$v/\sigma \sim$~0.1--0.3.
This conflicts with major merger simulations that predict 
rapid outer rotation: $v/\sigma \ga$~1 outside 2~$R_{\rm eff}$
(see Fig.~1, left; Weil \& Hernquist 1996; Bendo \& Barnes 2000).

The ideal comparison is to simulations of galaxy formation in the full cosmological context,
but treatments of key baryonic processes such as star formation and feedback are still maturing.
Current CDM models
do produce low specific angular momenta
similar to our observed ellipticals
(see Fig.~1, right).

\begin{figure}
\plottwo{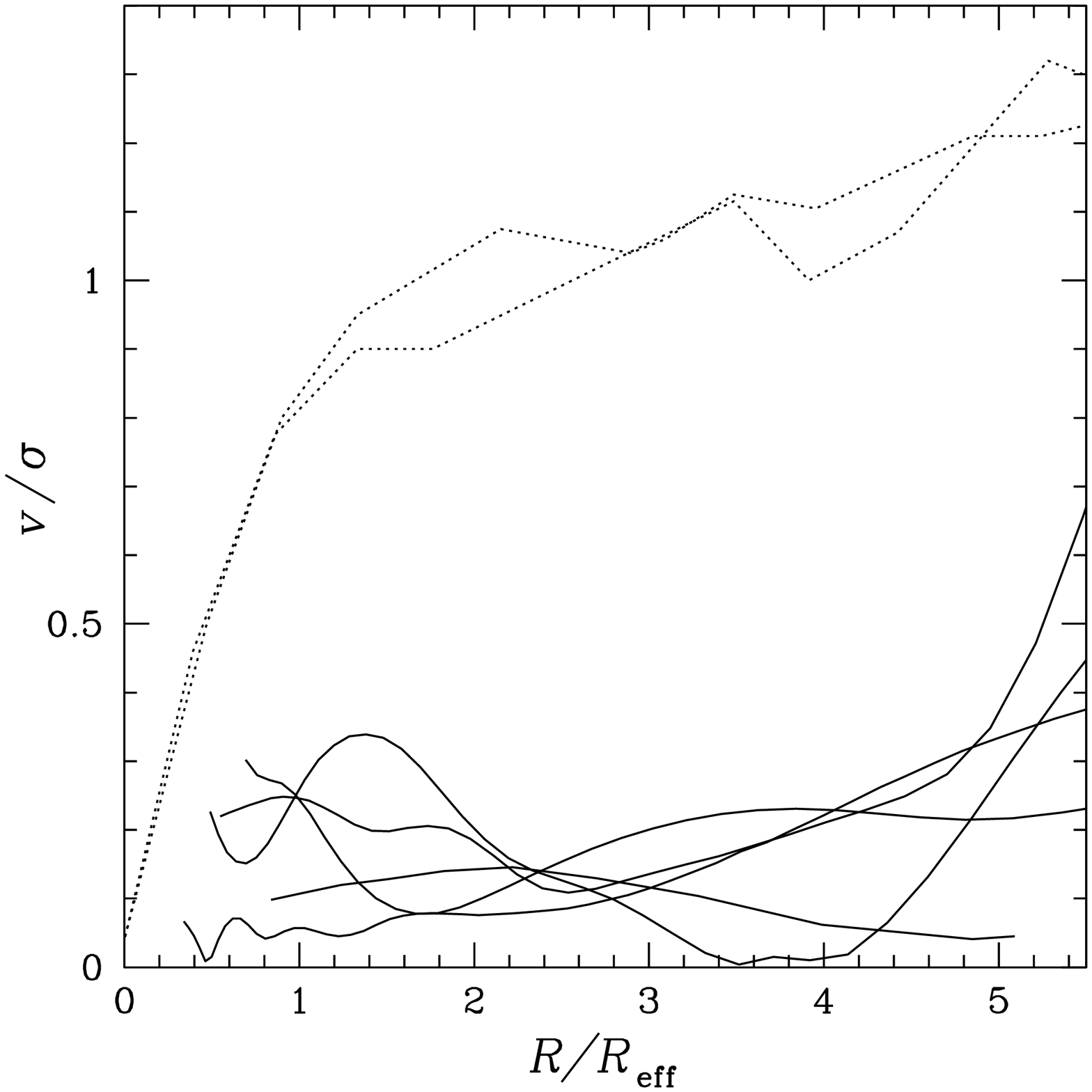}{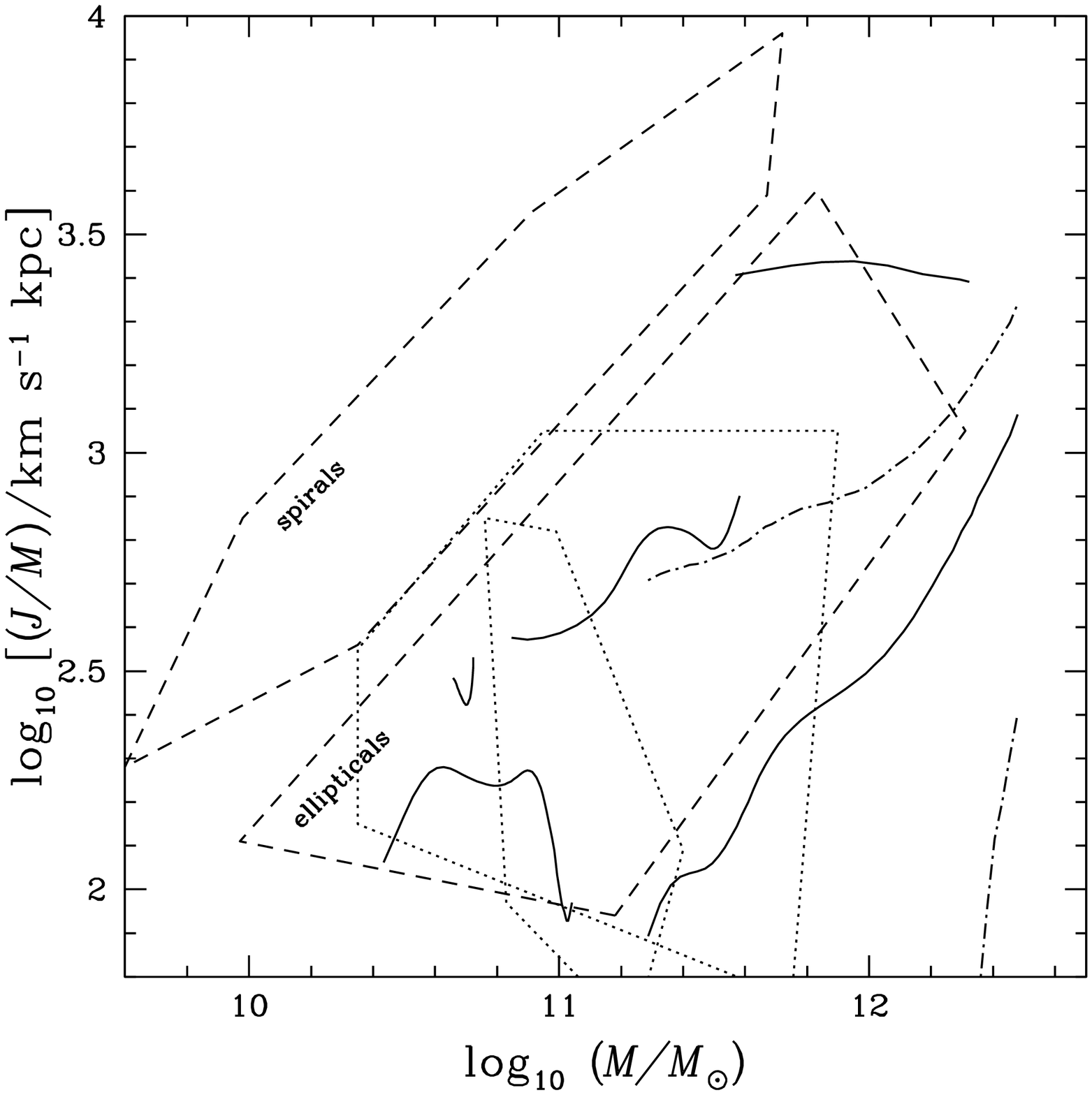}
\caption{
\emphasize{Left:} 
Rotational parameter as a function of projected radius
for five galaxies observed with PNe (solid lines),
and for a typical simulated merger remnant
(dotted lines; Weil \& Hernquist 1996).
\emphasize{Right:} 
Specific angular momentum as a function of enclosed mass, 
after Fall (1983).
Dashed lines outline observations for spiral disks and
elliptical's central parts.
Dotted lines show CDM hydrodynamical simulations for stellar components
(Navarro et al. 1995; Eke et al. 2000).
Solid lines show a run with radius for the PN galaxies,
where the 1-$\sigma$ uncertainties are illustrated for one case by dot-dashed lines.
}
\end{figure}

\section{Mass Content: skimpy halos}

Around M87 and M49, we find velocity dispersion profiles that rise slightly
with radius, suggesting massive extended dark halos,
as found by the GC and X-ray studies.
Combining all these constraints together will give us a detailed picture of
the mass and orbit distributions.
It is initially evident that the stellar orbit radial anisotropy increases into the halo.

The more ordinary ($\sim$$L^*$) ellipticals from our PN.S studies,
as well as NGC~4697 (M\'{e}ndez et al. 2001), show something entirely different:
their velocity dispersions decline rapidly with radius 
(see Fig. 2, left).
Simple Jeans models with a moderate degree of anisotropy indicate 
total masses consistent with the visible stars only:
a benchmark parameter $\Upsilon_{B\rm 5}$ ($M/L$ inside 5~$R_{\rm eff}$) is
$\sim$~6--15
(Romanowsky et al. 2003),
while stellar populations should have $\Upsilon_B \sim$~3--12 (Gerhard et al. 2001).
For NGC~3379, we have constructed more versatile orbit models
to allow for the infamous mass-anisotropy degeneracy,
and to extract as much information as possible out of the discrete PN velocity data
(Romanowsky \& Kochanek 2001).
These tightly constrain $\Upsilon_{B\rm 5}$ to be $7.1\pm0.6$ (Fig.~2, right).
There are still systematic uncertainties in this study,
notably the possibility that the galaxy contains a component that is flattened along the line-of-sight;
however, independent confirmation of a low $M/L$ comes from an HI ring (Schneider et al. 1989),
and from a steeply declining dispersion in the GCs (Beasley et al. 2004; Bergond et al. 2004)---which
are unlikely to reside in a flattened system.

\begin{figure}
\plottwo{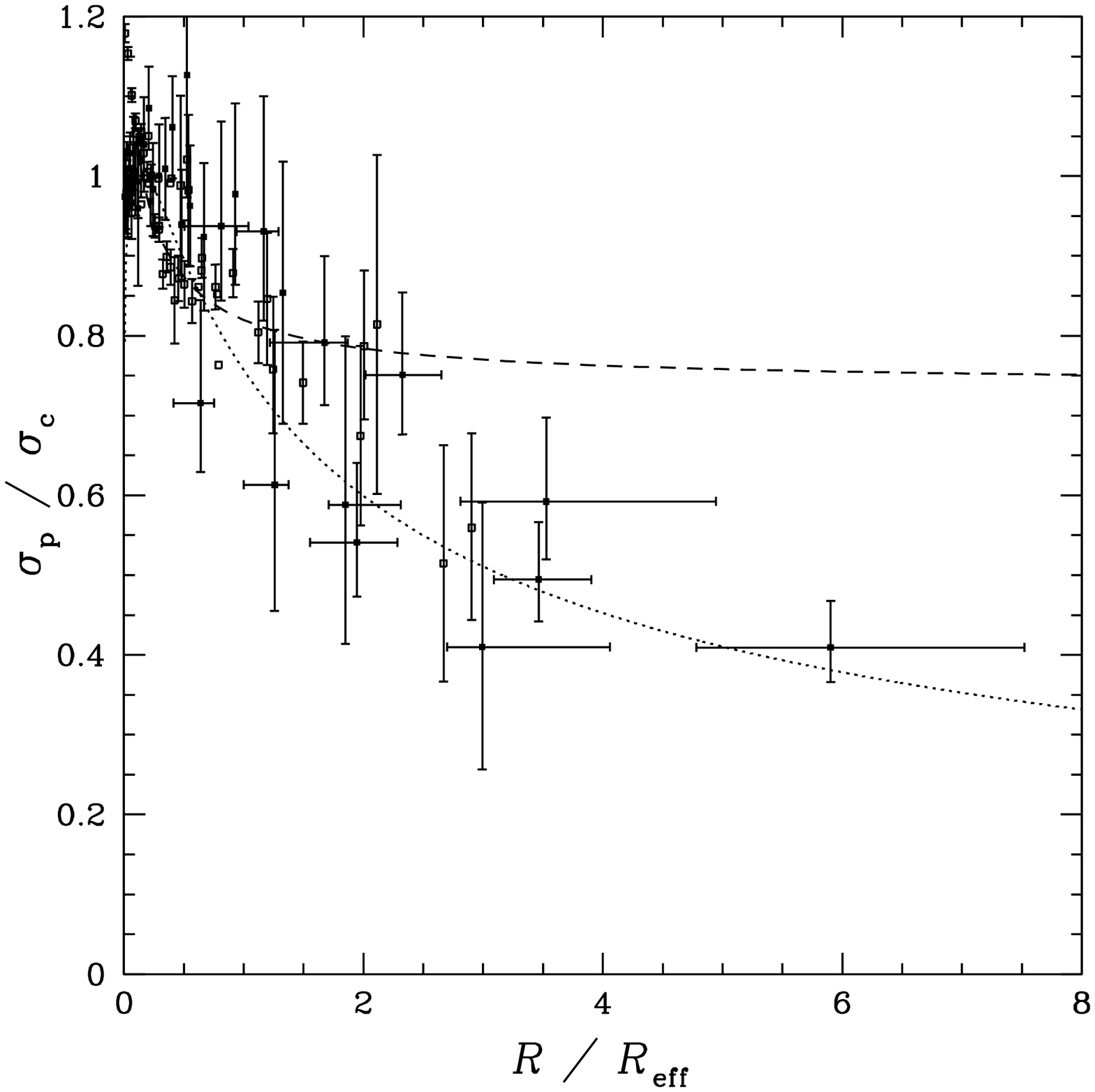}{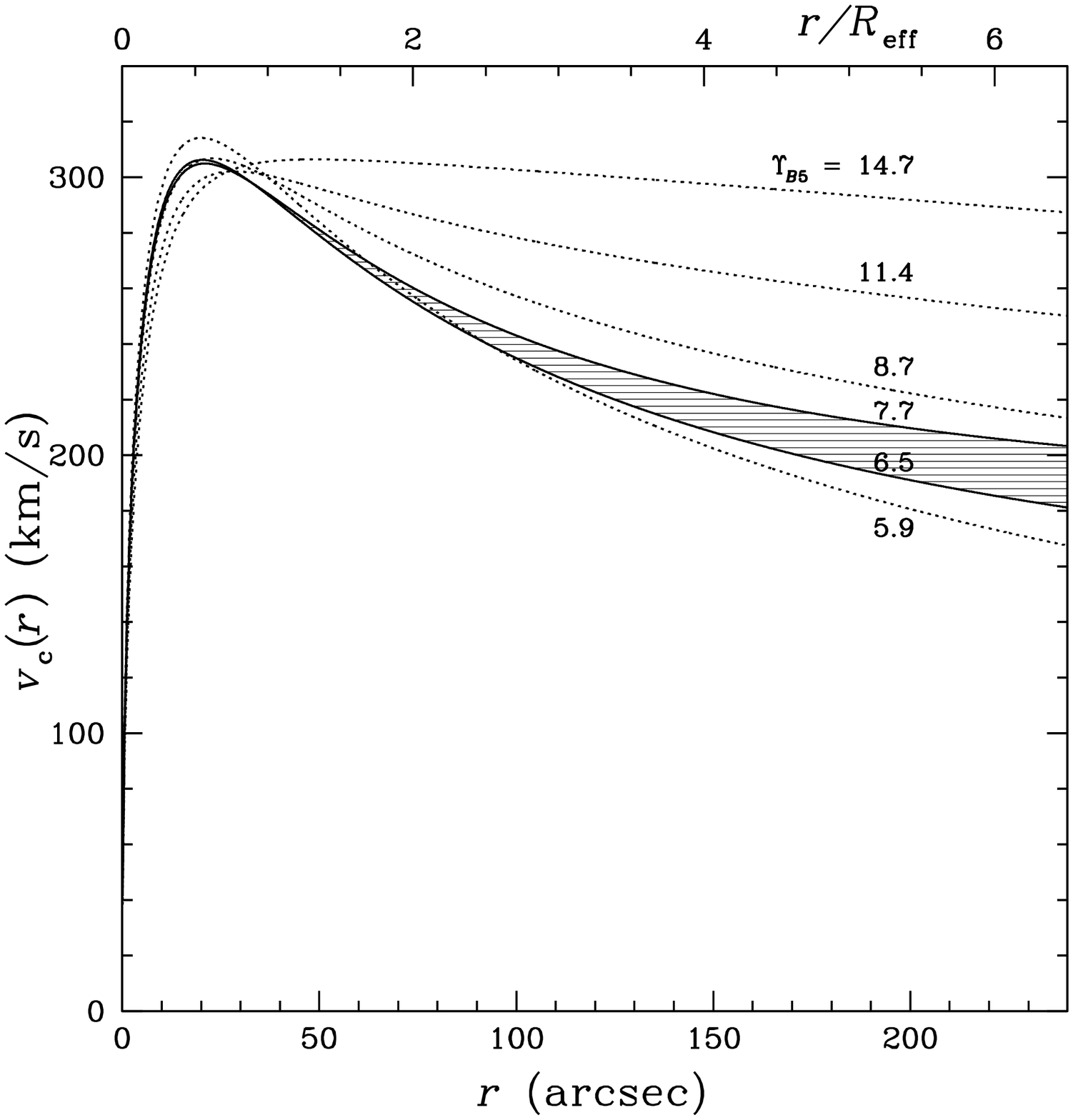}
\caption{
\emphasize{Left:}
 Projected velocity dispersion profiles for four elliptical galaxies
(see text)
scaled and stacked, as a function of radius, in units of $R_{\rm eff}$.
 Open points show PN data; solid points show long-slit stellar data.
 Simple model predictions are shown for comparison:
 a singular isothermal halo (dashed line) and a constant-$M/L$ galaxy (dotted line).
\emphasize{Right:}
Circular velocity profile of NGC~3379.
The solid lines and shaded area show the region permitted by orbit modeling
within the 68\% confidence limits.
Dotted lines show excluded models
(bottom: constant-$M/L$; top three: more dominant dark halos).
}
\end{figure}

For comparison, 
other
results from brighter galaxies typically show
$\Upsilon_{B\rm 5}\sim$~20--40 (Bahcall et al. 1995; Loewenstein \& White 1999).
Weak lensing estimates for $L^*$ ellipticals, extrapolated inwards in radius,
give $\Upsilon_{B\rm 5}\sim$~15--25
(Wilson et al. 2001; Seljak 2002).
Theoretical CDM predictions at these intermediate radii are not yet firm,
but
the best estimates
give $\Upsilon_{B\rm 5}\sim$~15--20
(e.g., Weinberg et al. 2004; Wright et al. 2004),
somewhat higher than we infer for the PN galaxies.

Thus, it seems there is a population of elliptical galaxies largely bereft of dark matter.
One explanation is that
their primordial CDM halos have
been stripped away or puffed up by interactions with other systems,
as can happen in galaxy clusters (Natarajan et al. 2002).
However, these ellipticals are not in such rich environments.
An alternative is that
this
is another manifestation of
the generic problem of
insufficiently centrally concentrated dark matter;
we infer halo concentrations $c\sim$~5 while CDM
predicts $c\sim$~15 \emphasize{before} including baryonic effects.
This interpretation is supported by studies at smaller radii with gravitational lenses---where
the dark matter fraction is lower than CDM expectations,
especially given additional priors on $H_0$
(Rusin et al. 2003)---and
with dynamical model fits to the fundamental plane (Borriello et al. 2003).
Further observational and theoretical investigations are needed to see
if this ``missing missing mass''
presents a major predicament for CDM.


\begin{references}
\reference {Baes, M., \& Dejonghe, H. 2001, \apj, 563, L19}
\reference {Bahcall, N. A., Lubin, L. M., \& Dorman, V. 1995, \apj, 447, L81}
\reference {Beasley, M. A., et al., in preparation}
\reference {Bendo, G. J., \& Barnes, J. E. 2000, \mnras, 316, 315}
\reference {Bergond, G., 
et al., in preparation}
\reference {Bertin, G., et al. 1994, \aap, 292, 381}
\reference {Borriello, A., Salucci, P., \& Danese, L. 2003, \mnras, 341, 1109}
\reference {Ciardullo, R., et al. 1998, \apj, 492, 62}
\reference {C\^{o}t\'{e}, P., et al. 2001, \apj, 559, 828}
\reference {C\^{o}t\'{e}, P., et al. 2003, \apj, 591, 850}
\reference {Douglas, N. G., et al. 2002, \pasp, 114, 1234}
\reference {Eke, V., Efstathiou, G., \& Wright, L. 2000, \mnras, 315, L18}
\reference {Fall, S. M. 1983, in Internal Kinematics and Dynamics of Galaxies, ed. E. Athanassoula (Dordrecht: Reidel), 391}
\reference {Gerhard, O., Kronawitter, A., Saglia, R. P., \& Bender, R. 2001, \aj, 121, 1936}
\reference {Lambas, D. G., Maddox, S. J., \& Loveday, J. 1992, \mnras, 258, 404}
\reference {Loewenstein, M., \& White, R. E., III. 1999, \apj, 518, 50}
\reference {Matsushita, K., et al. 2002, \aap, 386, 77}
\reference {M\'{e}ndez, R. H., et al. 2001, \apj, 563, 135}
\reference {Natarajan, P., Kneib, J.-P., \& Smail, I. 2002, \apj, 580, L11}
\reference {Navarro, J. F., Frenk, C. S., \& White, S. D. M. 1995, \mnras, 275, 56}
\reference {O'Sullivan, E., Ponman, T. J., \& Collins, R. S. 2003, \mnras, 340, 1375}
\reference {Peng, E. W., 
Ford, H. C., \& Freeman, K. C. 
2002, in 
Extragalactic Star Clusters,
ed. D. Geisler, 
E. K. Grebel, \& D. Minniti.
(San Francisco: ASP), 312}
\reference {Prada, F., et al. 2003, \apj, in press ({\tt astro-ph/0301360})}
\reference {Romanowsky, A. J., \& Kochanek, C. S. 2001, \apj, 553, 722}
\reference {Romanowsky, A. J., et al. 2003, Science, 301, 1696}
\reference {Rusin, D., Kochanek, C. S., \& Keeton, C. R. 2003, \apj, 595, 29}
\reference {Seljak, U. 2002, \mnras, 334, 797}
\reference {Schindler, S., Binggeli, B., \& B\"{o}hringer, H. 1999, \aap, 343, 420}
\reference {Schneider, S. E., et al., 1989, \aj, 97, 666}
\reference {Weil, M. L., \& Hernquist, L. 1996, \apj, 460, 101}
\reference {Weinberg, D. H., 
et al.
2004, \apj, in press ({\tt astro-ph/0212356})}
\reference {Wilson, G., Kaiser, N., Luppino, G. A., \& Cowie, L. L. 2001, \apj, 555, 572}
\reference {Wright, L. J.,
et al.,
2004, \mnras, submitted ({\tt astro-ph/0310513})}
\end{references}
\end{document}